\documentclass[12pt]{article}
\usepackage{epsfig}
\usepackage{amsmath}
\usepackage{hhline}
\usepackage{amssymb}
\usepackage{times}
\usepackage{cite}

\newlength{\dinwidth}
\newlength{\dinmargin}
\def\be{\begin{equation}}
\def\ee{\end{equation}}
\def\bea{\begin{eqnarray}}
\def\eea{\end{eqnarray}}
\def\etal{{\it et~al.}}

\begin{document}  
\newcommand{\pom}{{I\!\!P}}
\newcommand{\slowpi}{\pi_{\mathit{slow}}}
\newcommand{\fiidiii}{F_2^{D(3)}}
\newcommand{\fiidiiiarg}{\fiidiii\,(\beta,\,Q^2,\,x)}
\newcommand{\n}{1.19\pm 0.06 (stat.) \pm0.07 (syst.)}
\newcommand{\nz}{1.30\pm 0.08 (stat.)^{+0.08}_{-0.14} (syst.)}
\newcommand{\fiidiiiful}{F_2^{D(4)}\,(\beta,\,Q^2,\,x,\,t)}
\newcommand{\fiipom}{\tilde F_2^D}
\newcommand{\ALPHA}{1.10\pm0.03 (stat.) \pm0.04 (syst.)}
\newcommand{\ALPHAZ}{1.15\pm0.04 (stat.)^{+0.04}_{-0.07} (syst.)}
\newcommand{\fiipomarg}{\fiipom\,(\beta,\,Q^2)}
\newcommand{\pomflux}{f_{\pom / p}}
\newcommand{\nxpom}{1.19\pm 0.06 (stat.) \pm0.07 (syst.)}
\newcommand {\gapprox}
   {\raisebox{-0.7ex}{$\stackrel {\textstyle>}{\sim}$}}
\newcommand {\lapprox}
   {\raisebox{-0.7ex}{$\stackrel {\textstyle<}{\sim}$}}
\def\gsim{\,\lower.25ex\hbox{$\scriptstyle\sim$}\kern-1.30ex%
\raise 0.55ex\hbox{$\scriptstyle >$}\,}
\def\lsim{\,\lower.25ex\hbox{$\scriptstyle\sim$}\kern-1.30ex%
\raise 0.55ex\hbox{$\scriptstyle <$}\,}
\newcommand{\pomfluxarg}{f_{\pom / p}\,(x_\pom)}
\newcommand{\dsf}{\mbox{$F_2^{D(3)}$}}
\newcommand{\dsfva}{\mbox{$F_2^{D(3)}(\beta,Q^2,x_{I\!\!P})$}}
\newcommand{\dsfvb}{\mbox{$F_2^{D(3)}(\beta,Q^2,x)$}}
\newcommand{\dsfpom}{$F_2^{I\!\!P}$}
\newcommand{\gap}{\stackrel{>}{\sim}}
\newcommand{\lap}{\stackrel{<}{\sim}}
\newcommand{\fem}{$F_2^{em}$}
\newcommand{\tsnmp}{$\tilde{\sigma}_{NC}(e^{\mp})$}
\newcommand{\tsnm}{$\tilde{\sigma}_{NC}(e^-)$}
\newcommand{\tsnp}{$\tilde{\sigma}_{NC}(e^+)$}
\newcommand{\st}{$\star$}
\newcommand{\sst}{$\star \star$}
\newcommand{\ssst}{$\star \star \star$}
\newcommand{\sssst}{$\star \star \star \star$}

\newcommand{\tw}{\theta_W}
\newcommand{\sw}{\sin{\theta_W}}
\newcommand{\cw}{\cos{\theta_W}}
\newcommand{\sww}{\sin^2{\theta_W}}
\newcommand{\cww}{\cos^2{\theta_W}}
\newcommand{\trm}{m_{\perp}}
\newcommand{\trp}{p_{\perp}}
\newcommand{\trmm}{m_{\perp}^2}
\newcommand{\trpp}{p_{\perp}^2}
\newcommand{\alp}{\alpha_s}

\newcommand{\alps}{\alpha_s}
\newcommand{\sqrts}{$\sqrt{s}$}
\newcommand{\LO}{$O(\alpha_s^0)$}
\newcommand{\Oa}{$O(\alpha_s)$}
\newcommand{\Oaa}{$O(\alpha_s^2)$}
\newcommand{\PT}{p_{\perp}}
\newcommand{\JPSI}{J/\psi}
\newcommand{\sh}{\hat{s}}
\newcommand{\uh}{\hat{u}}
\newcommand{\MP}{m_{J/\psi}}
\newcommand{\PO}{I\!\!P}
\newcommand{\xbj}{x}
\newcommand{\xpom}{x_{\PO}}
\newcommand{\ttbs}{\char'134}
\newcommand{\xpomlo}{3\times10^{-4}}  
\newcommand{\xpomup}{0.05}  
\newcommand{\dgr}{^\circ}
\newcommand{\pbarnt}{\,\mbox{{\rm pb$^{-1}$}}}
\newcommand{\gev}{\,\mbox{GeV}}
\newcommand{\WBoson}{\mbox{$W$}}
\newcommand{\fbarn}{\,\mbox{{\rm fb}}}
\newcommand{\fbarnt}{\,\mbox{{\rm fb$^{-1}$}}}
%
%
\newcommand{\qsq}{\ensuremath{Q^2} }
\newcommand{\gevsq}{\ensuremath{\mathrm{GeV}^2} }
\newcommand{\et}{\ensuremath{E_t^*} }
\newcommand{\rap}{\ensuremath{\eta^*} }
\newcommand{\gp}{\ensuremath{\gamma^*}p }
\newcommand{\dsiget}{\ensuremath{{\rm d}\sigma_{ep}/{\rm d}E_t^*} }
\newcommand{\dsigrap}{\ensuremath{{\rm d}\sigma_{ep}/{\rm d}\eta^*} }

\begin{titlepage}

\noindent
DESY 01-100  \hfill  ISSN 0418-9833 \\
July 2001

\vspace*{3cm}

\begin{center}
    \begin{Large}

	{\bf Measurement of \boldmath$D^{*\pm}$ Meson Production and 
	$F_2^{c}$ in Deep-Inelastic Scattering at HERA}

	\vspace*{1cm}
	{\Large H1 Collaboration}

    \end{Large}
\end{center}

\vspace*{3cm}
\def\gev{\rm GeV}
\def\ie{\it i.e.}
\def\etal{\hbox{$\it et~al.$}}
\def\clb#1 {(#1 Coll.),}
\hyphenation{do-mi-nant}

\begin{abstract}
\noindent
The inclusive production of $D^{*\pm}(2010)$ mesons in deep-inelastic 
scattering
is studied with the H1 detector at HERA.
In the kinematic region
$1<Q^2<100$~GeV$^2$ and
$0.05<y<0.7$
an $e^+p$~cross section for inclusive $D^{*\pm}$ meson production of 
$8.50\pm 0.42\,$(stat.)$^{+1.21}_{-1.00}\,$(syst.)$\,$nb
is measured in the visible range
$p_{t\,D^*}~>~1.5$~GeV and
$|\eta_{D^*}|~<~1.5$. 
Single and double differential inclusive $D^{*\pm}$ meson cross sections are 
compared to 
perturbative QCD calculations in
two different evolution schemes. 
 The charm contribution to the proton structure, 
$F_2^{c}\left(x,Q^2\right)$, is determined by extrapolating the 
visible charm cross section to the full phase space.  
This contribution 
is found
to rise from about 10\% at $Q^2=1.5~\gev^2$ to more than 25\% at 
$Q^2=60~\gev^2$ corresponding to $x$ values ranging from $5\cdot 10^{-5}$ to 
$3\cdot 10^{-3}$.
\end{abstract}
\vspace*{1.5cm}
\begin{center}
    {\it{Submitted to Phys. Lett. B}}
\end{center}
\end{titlepage}

\noindent
C.~Adloff$^{33}$,              
V.~Andreev$^{24}$,             
B.~Andrieu$^{27}$,             
T.~Anthonis$^{4}$,             
V.~Arkadov$^{35}$,             
A.~Astvatsatourov$^{35}$,      
A.~Babaev$^{23}$,              
J.~B\"ahr$^{35}$,              
P.~Baranov$^{24}$,             
E.~Barrelet$^{28}$
% ...

\bigskip{\noindent\it
 $ ^{1}$ I. Physikalisches Institut der RWTH, Aachen, Germany$^{ a}$ \\
 $ ^{2}$ III. Physikalisches Institut der RWTH, Aachen, Germany$^{ a}$ \\
% ...
$ ^{4}$ Inter-University Institute for High Energies ULB-VUB, Brussels;
          Universitaire Instelling Antwerpen, Wilrijk; Belgium$^{ c}$ \\
% ...
$ ^{21}$ Physics Department, University of Manchester,
          Manchester, UK$^{ b}$ \\
% ...
$ ^{23}$ Institute for Theoretical and Experimental Physics,
          Moscow, Russia \\
 $ ^{24}$ Lebedev Physical Institute, Moscow, Russia$^{ e,h}$ \\
% ...
$ ^{27}$ LPNHE, Ecole Polytechnique, IN2P3-CNRS, Palaiseau, France \\
 $ ^{28}$ LPNHE, Universit\'{e}s Paris VI and VII, IN2P3-CNRS,
          Paris, France \\
% ...
$ ^{33}$ Fachbereich Physik, Bergische Universit\"at Gesamthochschule
          Wuppertal, Wuppertal, Germany$^{ a}$ \\
% ...
$ ^{35}$ DESY, Zeuthen, Germany$^{ a}$ \\
% ...
$ ^a$ Supported by the Bundesministerium f\"ur Bildung, Wissenschaft,
      Forschung und Technologie, FRG,
      under contract numbers 7AC17P, 7AC47P, 7DO55P, 7HH17I, 7HH27P,
      7HD17P, 7HD27P, 7KI17I, 6MP17I and 7WT87P \\
 $ ^b$ Supported by the UK Particle Physics and Astronomy Research
      Council, and formerly by the UK Science and Engineering Research
      Council \\
 $ ^c$ Supported by FNRS-NFWO, IISN-IIKW \\
% ...
$ ^e$ Supported by the Deutsche Forschungsgemeinschaft \\
% ...
$ ^h$ Supported by Russian Foundation for Basic Research
      grant no. 96-02-00019 \\
% ...
}

\newpage

\section{Introduction}
Results on inclusive $D^{*\pm}$ meson 
 production in deep-inelastic $ep$ scattering (DIS) and on the charm 
contribution to the proton structure function, $F_2^{c}$, at HERA 
have been published by the H1 and the ZEUS collaborations 
\cite{h1f2c,zeusf2c,zeusf2c1}.
These data, together with earlier fixed target data \cite{emc},
 have shown clear evidence that the dynamics
of charm production in $ep$ scattering is described by the photon gluon fusion
process, 
which is 
sensitive to the gluon density in the proton \cite{h1gluon} and allows 
its universality to be tested.
  
Early results on $F^c_2$ from the H1 experiment \cite{h1f2c}
were based on an integrated luminosity
of 3~pb$^{-1}$ collected during the 1994 HERA running
and were therefore statistically limited.
The current analysis uses data from the 1996 and 1997 HERA running periods,
yielding a significantly larger integrated luminosity of 18.6~pb$^{-1}$.
Furthermore, the improved instrumentation in the backward region 
of the H1 detector enables the kinematic range in four-momentum transfer 
squared to the virtual photon, $Q^2$, 
to be significantly extended down to $1~\gev^2$.
Hence, more precise tests of perturbative QCD (pQCD) become possible.

This paper is organized as follows: a short discussion of the different 
approaches to open charm production in perturbative QCD calculations 
is followed by a
description of  the experimental set-up and details of the analysis; the 
inclusive cross sections for $D^{*\pm}$ meson production are then presented 
and compared to QCD predictions. Finally, they are used to derive the 
charm contribution to the proton structure function, $F_2^{c}$.

\section{Models of Open Charm Production}
\label{dglap}
The description of open heavy flavour production in electron proton collisions
is based on perturbative QCD.
In leading order (LO), the photon gluon fusion process  
($\gamma g\rightarrow Q\overline Q$) is the dominant contribution \cite{h1f2c}.
Next-to-leading order (NLO)
calculations in several schemes are available 
\cite{riemersma,riemersma2,harris,acot,collins}.
All approaches assume that $Q^2$ and the heavy quark mass $m_Q$ provide a hard
enough scale to allow the applicability of pQCD and to 
guarantee the validity of the factorization theorem.

Here, the ``massive approach'' is adopted, i.e. a fixed order 
calculation with massive quarks assuming three active flavours in the
proton. The momentum densities of the three light quarks and the gluon in the 
proton
are evolved by the DGLAP equation \cite{dglapref}. 
The heavy quarks are assumed to be
produced only at the 
perturbative level \cite{riemersma} via photon gluon fusion. 
These calculations are considered reliable in the regime  $Q^2\approx m_Q^2$. 
However, they break down at some scale $Q^2\gg m_Q^2$ due to large logarithms 
$\sim\ln(Q^2/m_Q^2)$. 

Based on fixed order $\alpha_s^2$ calculations in the coefficient functions 
\cite{riemersma} programs for different applications were developed. 
The Riemersma et al. program \cite{riemersma2} can be used to calculate
inclusive quantities of heavy quark production, like $F_2^c(x,Q^2)$, while the 
HVQDIS program \cite{harris,hvqdis} allows the calculation of
exclusive quantities by providing the four-momenta of 
the outgoing partons.
In the version of the program used here
charmed quarks are fragmented in the photon~-~proton centre of mass frame
into $D^{*\pm}$
mesons using the Peterson fragmentation function \cite{peterson}, which 
is controlled by a single parameter $\epsilon_c$. In addition, to account for 
the experimentally observed $p_t$ smearing of hadrons with respect to the 
quark direction, the $D^{*\pm}$ meson has been given a transverse momentum 
$p_t$ with respect to the charm quark, according to the function
$p_t\cdot\exp(-\alpha p_t)$.
The parameter $\alpha$ is chosen such that an average transverse momentum 
$\langle p_t\rangle\approx 350$~MeV is obtained as observed in $e^+e^-$ data
\cite{epem}.
With this procedure it becomes possible to calculate differential inclusive 
$D^{*\pm}$ meson cross sections in the experimentally visible phase space 
region.

The CCFM evolution equation \cite{ccfm} is expected to be more appropriate 
to describe the parton evolution at small $x$. In the parton cascade, 
gluons are emitted in an angular ordered manner to account for coherence
effects. Due to this angular ordering, the gluon distribution 
depends on the maximum allowed angle in addition to the 
momentum fraction $x$ and the transverse momentum of the propagator gluon.
The cross section is then calculated according to the $k_t$-factorization
theorem by convoluting the unintegrated gluon density with the off-shell 
photon gluon fusion matrix element with massive quarks for the hard 
scattering process.  

It has been shown previously \cite{f2cccfm} that $F_2$ and $F_2^{c}$
can be reasonably well described within the CCFM framework.  
In addition a solution of the CCFM 
equation has been obtained recently \cite{cascade} from a fit to $F_2$
which is able to describe the cross section for forward jet production,
where significant differences to the expectation in the DGLAP evolution 
scheme are seen. 
Using this solution the hadron level Monte Carlo generator
CASCADE has been developed \cite{hannes}. This allows the full generation of 
charm events 
including the initial state gluon radiation according to the
CCFM equation and the fragmentation of partons by the Lund String model.
The fragmentation of charmed quarks to $D^{*\pm}$ mesons is performed 
using the Peterson fragmentation function. 

\section{Detector and Simulation}
The data have been collected with the H1 detector \cite{h1det} at HERA
during the running periods of 1996 and 1997 when
HERA operated with 27.5 GeV positrons and 820~GeV protons colliding 
at a centre of mass energy of $\sqrt{s}=300$~GeV.
The following detector components are important for this analysis.
 The scattered positron is 
identified and measured in the SpaCal \cite{spacal}, a lead-scintillating 
fibre calorimeter situated in the backward region\footnote{ The positive 
$z$-axis of the H1 reference frame, which defines the forward direction,
is given by the
outgoing proton direction.} of the H1 detector. The SpaCal also provides 
time-of-flight information for trigger purposes. A four double-layer 
backward drift chamber (BDC) \cite{bdc} is mounted in front of the SpaCal 
in order to improve the angular measurement of the scattered positron.
Charged particle tracks are reconstructed by 
two cylindrical central jet drift chambers (CJC) \cite{h1det,cjc} placed 
concentrically around the beam-line in a homogeneous magnetic field of 
1.15 Tesla. The CJC  also provides trigger information \cite{dcrphi} 
based on the detection of track segments.
Double layers of cylindrical multi-wire proportional chambers (MWPC) 
\cite{mwpc} for triggering purposes are positioned inside and in-between 
the two jet chambers. 
The luminosity is determined from the rate of the Bethe-Heitler reaction
$ep\rightarrow ep\gamma$. 

Monte Carlo simulation programs
 are used to simulate detector effects and  to estimate
the systematic uncertainties associated with the measurement.
For the determination of the acceptance of the detector and the 
$D^{*\pm}$ selection efficiencies, heavy flavour (charm and bottom) DIS events 
are generated using the AROMA 2.2 \cite{aroma} program. 
This program, which is based on the DGLAP evolution scheme,
simulates neutral current heavy quark production via photon gluon 
fusion in leading order QCD including parton showers and heavy quark mass 
effects. The mass of the charm quark is
chosen to be $m_c=1.5\;\gev$ while the factorization and renormalization 
scales are set to $\mu=\sqrt{\hat s}$, 
where $\hat s$ denotes the square of the invariant mass of the heavy quark 
system.
The GRV94-LO \cite{grvlo} parton density functions (PDF's) are used for the 
proton. Hadronization is performed in the 
Lund String Model \cite{lundf}, as implemented in JETSET 7.4 \cite{jetset}. 
The momentum fraction of the charm quark
carried by the $D^{*\pm}$ meson is determined according to the Peterson model
\cite{peterson} with the fragmentation parameter $\epsilon_c=0.078$ 
\cite{argus}. The influence of the details of the fragmentation process on
the acceptances and efficiencies has been investigated by (a) varying the 
Peterson fragmentation parameter between $\epsilon_c=0.035$, 
as favoured in \cite{oleari},  and $\epsilon_c=0.1$ which seems to yield a 
better description of the hadronic final state in $D^{*\pm}$ events, (b) 
applying the symmetric Lund fragmentation function \cite{lundf} also to the 
$D^{*\pm}$ mesons and (c) using the HERWIG \cite{herwig} program which is 
based on the cluster hadronization model \cite{webber}.
The inaccuracy due to the uncertainty in the QCD parameters
is studied by varying the 
charm quark mass $m_c$ and by changing 
the factorization and renormalization scales to $\mu=\sqrt{Q^2+4m_c^2}$. The 
dependence of the 
acceptances and efficiencies on the QCD evolution scheme has been
determined also by using the CASCADE \cite{cascade} event generator. Finally,
the influence of QED radiation on the efficiency is determined using
the RAPGAP \cite{rapgap} program interfaced to HERACLES 4.1 \cite{herakles}.   
All Monte Carlo generated events are fed into the GEANT \cite{geant} based 
simulation of the H1 detector and are subjected to the 
same reconstruction and analysis chain as used for the data.

\section{Kinematics\label{kine}}
This analysis is restricted to those DIS events which have  a
scattered positron detected in the backward region of the detector.
At fixed center of mass energy $\sqrt{s}$ the
kinematics of the inclusive scattering process $ep\rightarrow eX$ can be
completely determined by any two of the independent Lorentz invariant 
variables: the Bjorken scaling variable $x$, the lepton inelasticity $y$, 
the four-momentum squared $Q^2=-q^2$ of the virtual photon and the invariant mass squared
$W^2$ of the hadronic final state. In this analysis, these variables are 
determined from the measurement of the energy $E^\prime_e$ and the polar 
angle $\Theta_e$ 
of the scattered positron according to the expressions
\begin{equation}
\begin{array}{ccc}\displaystyle
Q^2=4E_eE^\prime_e\cos^2\left(
\frac{\Theta_e}{2}\right)&\quad\quad&\displaystyle
y=1-\frac{ E^\prime_e}{ E_e}
\sin^2\left(\frac{ \Theta_e}{2}\right)
\cr\cr\displaystyle
x=\frac{ Q^2}{ ys}&\quad\quad&\displaystyle
W^2=Q^2\left(\frac{ 1-x}{ x}\right)\cr
\end{array}
\end{equation}
where $s=4E_eE_p$ and $E_e$ and $E_p$ denote the energies of the incoming
positron and proton, respectively 
(the positron and proton masses are neglected). 

\section{Event Selection}

The events for this analysis were triggered by a coincidence of 
an electromagnetic cluster in the SpaCal with a charged track signal from the 
CJC and a vertex which is coarsely reconstructed from the MWPC information.
The positron is identified as the most energetic cluster with 
$E_e^\prime>8$~GeV as described in \cite{h1nlo2000}. 
The cluster radius\footnote{The cluster radius is defined as 
$\sum_i\log(E_i)\cdot d_i/\sum_i\log(E_i)$ 
where the sum runs over all cells in the
cluster: $E_i$ is the normalized energy of the cell $i$ and $d_i$ 
is the distance of the cell $i$ from the cluster centre of gravity.} 
is required to be less than 4~cm, consistent with an
electromagnetic energy deposition, and the cluster center of gravity is 
required to be within 1.5~cm of the extrapolation of a 
charged track segment from the backward drift chamber BDC. 
The geometrical acceptance of the SpaCal and BDC 
imposes a limitation 
on the positron scattering angle of 
$\Theta_e<177.5^\circ$.
These limits and requirements
 restrict the accessible range in $Q^2$ from 1~GeV$^2$ to 100~GeV$^2$ and 
in the lepton inelasticity to $y<0.7$. 
To measure the event kinematic quantities with sufficiently good resolution 
$y$ is further constrained to $y>0.05$. 
Good agreement is observed for all quantities related to the
scattered positron
between data and the prediction of the AROMA Monte Carlo simulation.  
The contribution due to photoproduction background, i.e. $Q^2<1$~GeV$^2$,
is everywhere smaller than 1\% in the selected kinematic region.

\begin{figure}[htb] 
    \begin{center}
    \epsfig{file=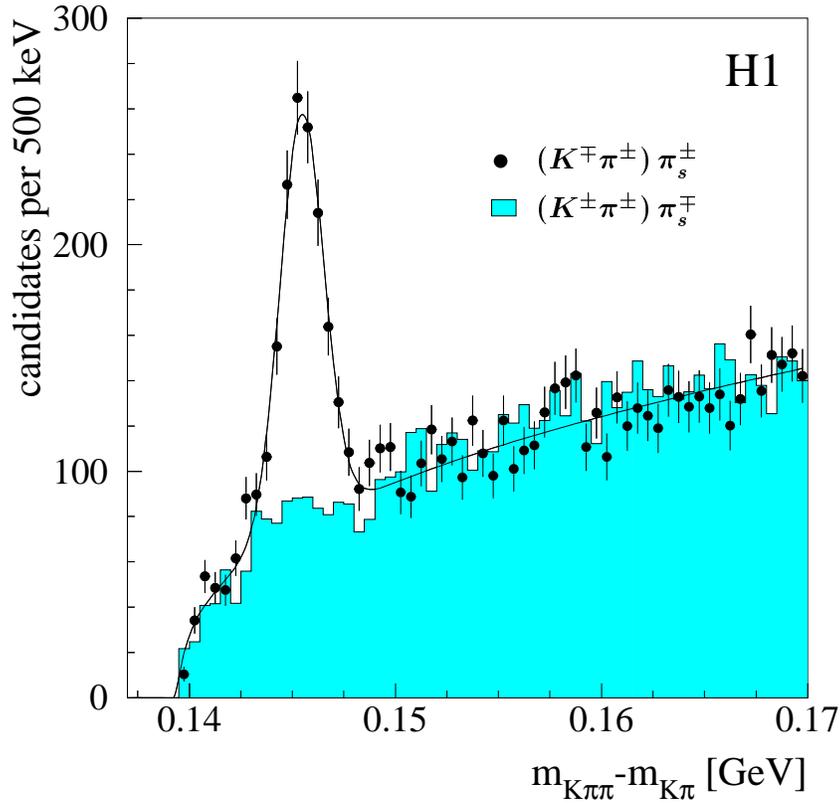,width=0.85\textwidth}
\caption{\label{fig1}{Distribution of the mass difference 
$\Delta{m} = m(K^\mp \pi^\pm \pi_s^\pm) - m(K^\mp \pi^\pm)$ for 
DIS events with $D^0$ candidates in the visible range $|\eta_{K\pi\pi}|<1.5$
and  $p_{t\,K\pi\pi}>1.5$~GeV.
The data points are obtained from the $K^\mp \pi^\pm$ mass combinations
fulfilling $|m(K^\mp \pi^\pm) - m_{D^0}| <$ 70 MeV.
The solid line represents the result of the fit described in the text.
The shaded histogram shows the background expectation from the like sign 
$K^\pm\pi^\pm$
pairs.
}}
\end{center}
\end{figure}

Charm production is identified by the reconstruction of
$D^{*\pm}$ mesons in the decay chain  
\begin{equation}
D^{*+}\rightarrow D^0\pi_{slow}^+\rightarrow (K^-\pi^+)\pi_{slow}^+ ~(+c.c.) 
\end{equation}   
using the $D^{*}-D^0$ mass difference method
\cite{feldmann}. 
The decay products are detected in the central track detector.
For each accepted track 
particle identification is applied using the measurement of the energy loss, 
${\rm d}E/{\rm d}x$, in the central track detector. 
In order to reconstruct a $D^{*\pm}$ meson candidate, unlike-sign charged 
tracks are first combined to 
form $K^\mp\pi^\pm$ pairs in which one of the particles should be consistent 
with a kaon and the other with
a pion according to their d$E/$d$x$ measurements.
Among all possible oppositely charged $K^\mp\pi^\pm$ pairs,
those with an invariant mass consistent within $\pm$70~MeV  of the $D^0$
mass are combined 
with a track of a second pion candidate (``$\pi^\pm_{slow}$'') having a charge
opposite in sign to that of the kaon.  
In  Fig. \ref{fig1} a clear peak is observed in the distribution of the 
mass difference $\Delta m=m_{K\pi\pi}-m_{K\pi}$ 
around the nominal $D^{*\pm}-D^0$ mass difference of 145.4 MeV. 
A fit to this distribution
using a Gaussian for the signal and a term $(\Delta m-m_\pi)^\alpha$ 
for the background  yields 
a total of $973\pm40$ $D^{*\pm}$ mesons in the acceptance 
range of pseudorapidity\footnote{The pseudorapidity of a particle is defined as
$\eta\equiv-\ln\tan(\Theta/2)$.} $|\eta_{K\pi\pi}|<1.5$ 
and transverse momentum $p_{t\,K\pi\pi}>1.5$~GeV.

\section {Inclusive Cross Sections}
The integrated and differential Born level cross sections for 
$D^{*\pm}$ meson production in DIS
 are calculated from the observed number $ N_{D^{*\pm}}$ 
of  $D^{*\pm}$ candidates, according to 
\begin{equation}
\sigma_{\rm{vis}}(e^+p \rightarrow e^+D^{*\pm} X)=
\frac{N_{D^{*\pm}}~(1-r)}
{ {\cal L}_{int} \cdot B \cdot\epsilon\cdot(1+\delta_{rad})  }~.
\label{sigman}
\end{equation}
Here, $r$ stands for the contribution of reflections in the $D^0$ mass 
window, coming from $D^0$ channels other than the one studied 
in this analysis. The value of $r$ amounts on average to about 0.03. 
The integrated luminosity
is denoted by 
${\cal L}_{int}$ while $B$ refers to the branching ratio 
$B = B(D^{*+} \rightarrow D^0 \pi^+) \cdot B(D^0 \rightarrow K^- \pi^+) =
0.0259\pm0.0006$ \cite{PDG}. The detection efficiency $\epsilon$ is estimated 
to be 22.5\% using AROMA. 
The radiative correction $\delta_{rad}$ which correct to the
single photon exchange cross sections are obtained from the program HECTOR 
\cite{hector}. Depending on the kinematic region $\delta_{rad}$ varies from
+0.11 at small $x$ and $Q^2$ to -0.02 at large $Q^2$.
For the integrated visible cross section it averages
to 0.03.

\subsection{Integrated Cross Section}
The inclusive cross section for $D^{*\pm}$ meson production  
in the kinematic region $1<Q^2<100\gev^2$ and $ 0.05\,<\,y\,<\,0.7$, and 
in the visible $D^{*\pm}$ range $|\eta_{D^*}|<1.5$ and $p_{t\,D^*}>1.5$~GeV
is found to be
$$\sigma_{vis}(e^+p \rightarrow e^+D^{*\pm} X)=
8.50\pm 0.42(\rm{stat.})^{+1.02}_{-0.76}\,
(\rm{syst.})\pm 0.65(\rm{model})\,\rm{nb}.$$
The errors refer to those from statistics, experimental systematics and
additional systematics related to the changes in efficiency obtained by using 
different Monte Carlo generators and varying the model parameters. 

The experimental systematic uncertainties are summarized in Table~\ref{expsys}.
The largest contribution is due to the uncertainty in the track reconstruction 
efficiency. Other important sources include uncertainties in the extraction 
of the $D^{*\pm}$ signal, i.e. the determination of the background shape in 
the $\Delta m$ distribution and the $D^0$ mass resolution. 
The uncertainties due to model dependencies, as summarized in 
Table~\ref{modelsys}, include the incomplete understanding of the 
fragmentation process, the uncertainty due to the charm quark mass, the 
sensitivity to the factorization and renormalization scales and the change 
of acceptance due to QED effects at the positron vertex. 
The largest effect on the efficiency is observed by 
changing the charm quark mass from $m_c=1.5~\gev$ 
in the reference Monte Carlo dataset to $m_c=1.3~\gev$ 
and by changing the fragmentation models and their parameters.

The visible inclusive $D^{*\pm}$ meson production cross section has been 
calculated in the NLO DGLAP scheme with the HVQDIS program using the 
GRV98-HO parton densities in the proton \cite{grv}. The predictions range
from 5.17~nb for a charm quark mass $m_c=1.5~\gev$ and Peterson fragmentation 
parameter $\epsilon_c=0.10$ to 7.02~nb for $m_c=1.3~\gev$ and 
$\epsilon_c=0.035$. The hadronization fraction 
$f(c\rightarrow D^{*+})=0.233\pm0.010\pm0.011$ \cite{hadro} has been used.
For the same variation of $m_c$ and $\epsilon_c$,
calculations based on the CCFM evolution, as implemented in
the CASCADE program, yield a significantly higher cross section of
8.04~nb and 10.77~nb, respectively.

Disregarding the small differences in the kinematic range, good agreement is
observed in the inclusive $D^{*\pm}$ meson production cross section 
with the result obtained by the ZEUS experiment \cite{zeusf2c1}.
The measured value of this cross section agrees better with the CASCADE 
prediction than with that from HVQDIS.
In previous publications \cite{h1f2c,h1gluon}
H1 reported much better agreement between data and predictions from
the HVQDIS program. The larger difference obtained now is due to the new 
determination of the charm quark hadronization fraction 
$f(c\rightarrow D^{*+})$ which is 16\% smaller than the previous value.

%
%
\subsection{Differential Cross Sections\label{difwq}}
\begin{figure}[htb] 
\begin{picture}(15.,19.)
\put(1,12.5){\epsfig{file=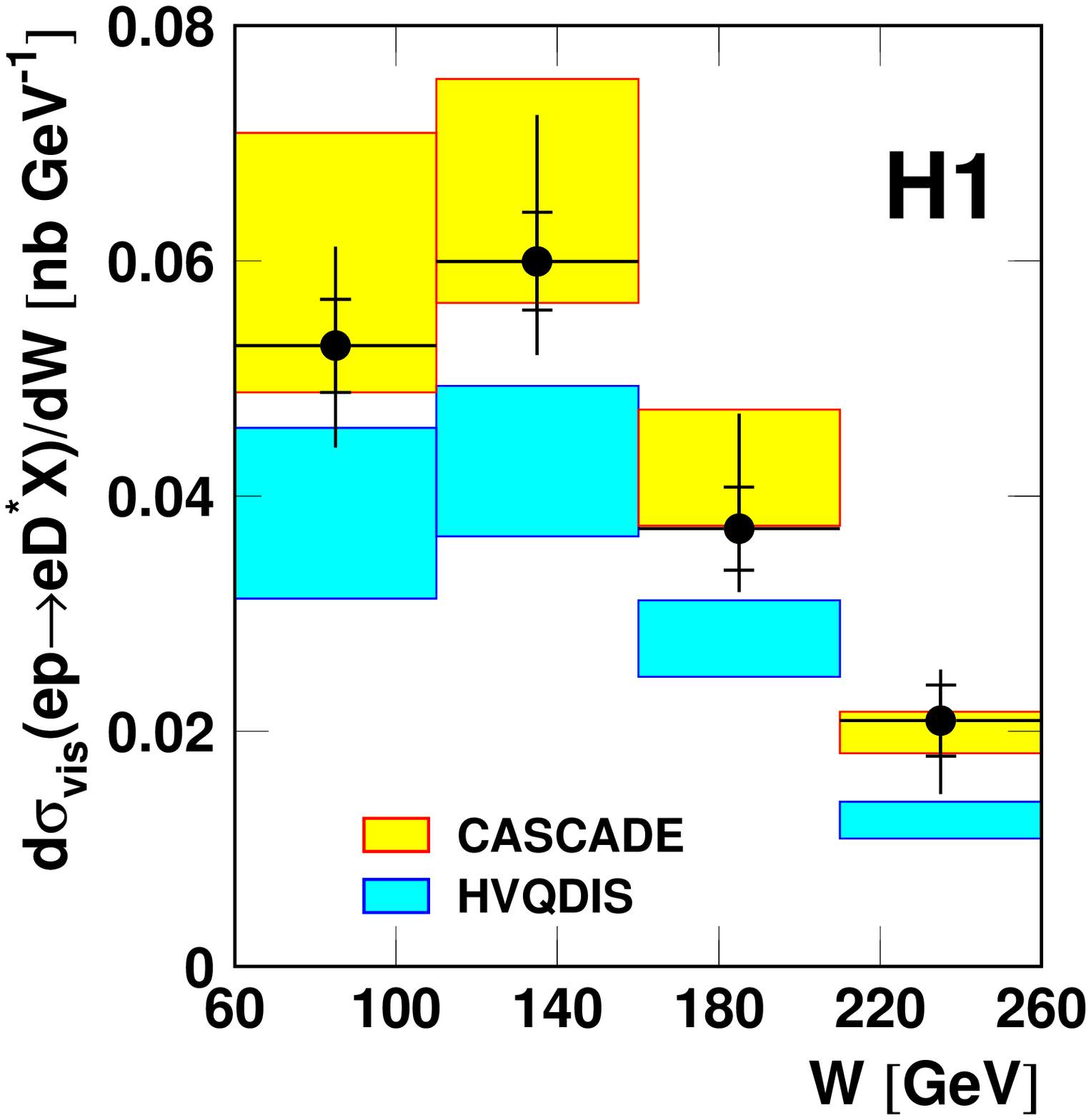,width=7.5cm}}
\put(8,12.5){\epsfig{file=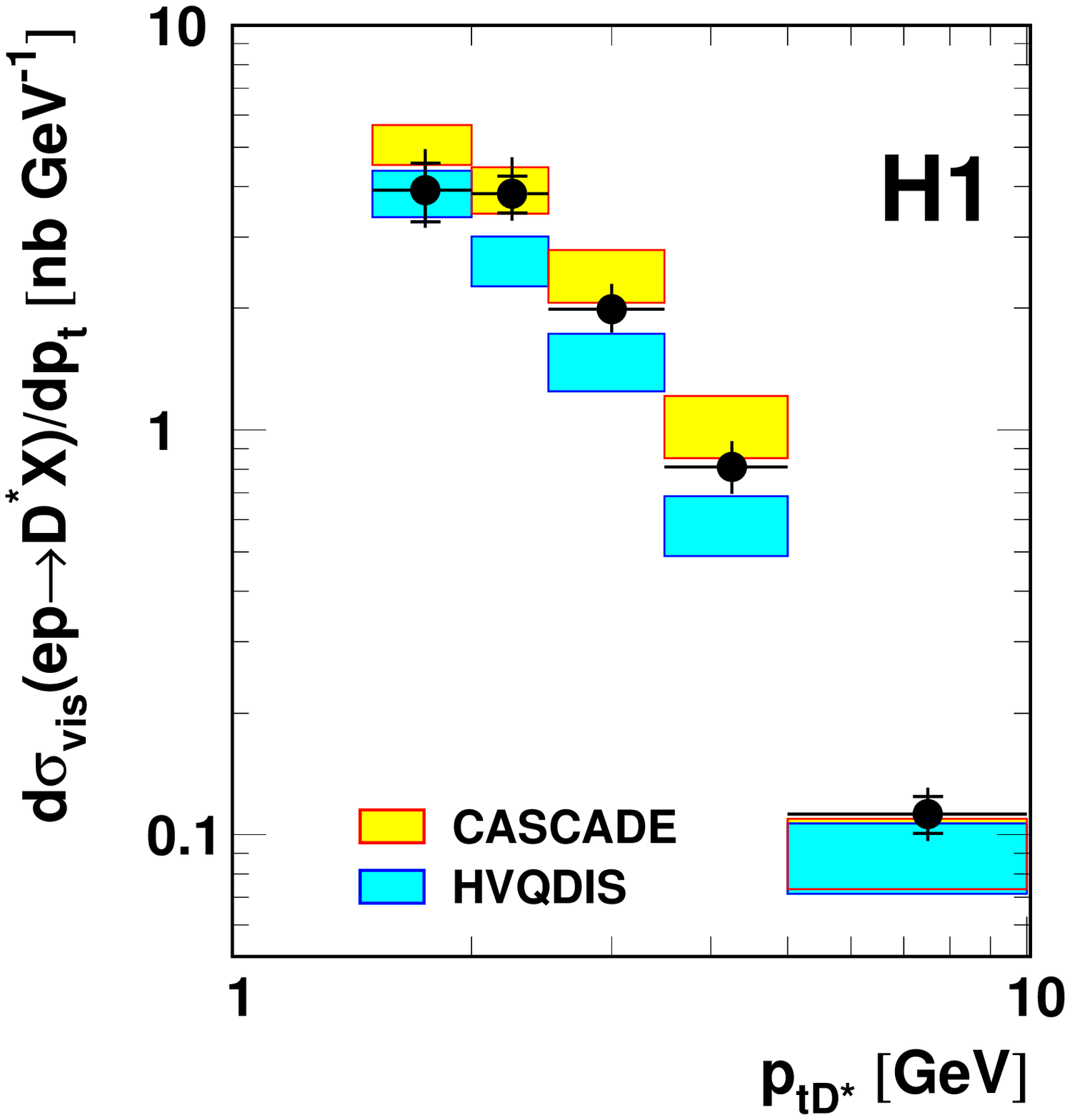,width=7.5cm}}
\put(1,5.75){\epsfig{file=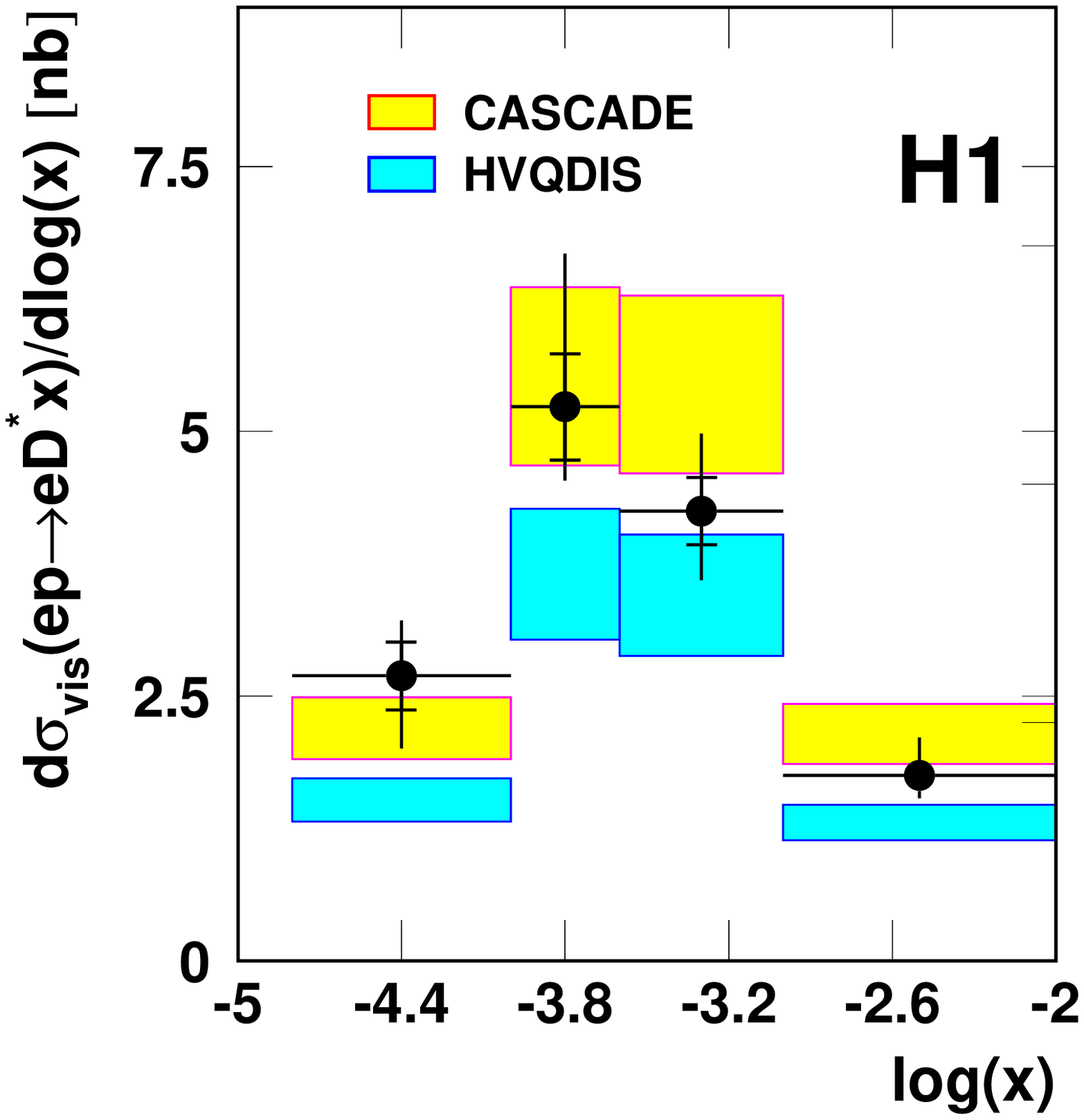,width=7.5cm}}
\put(8,5.75){\epsfig{file=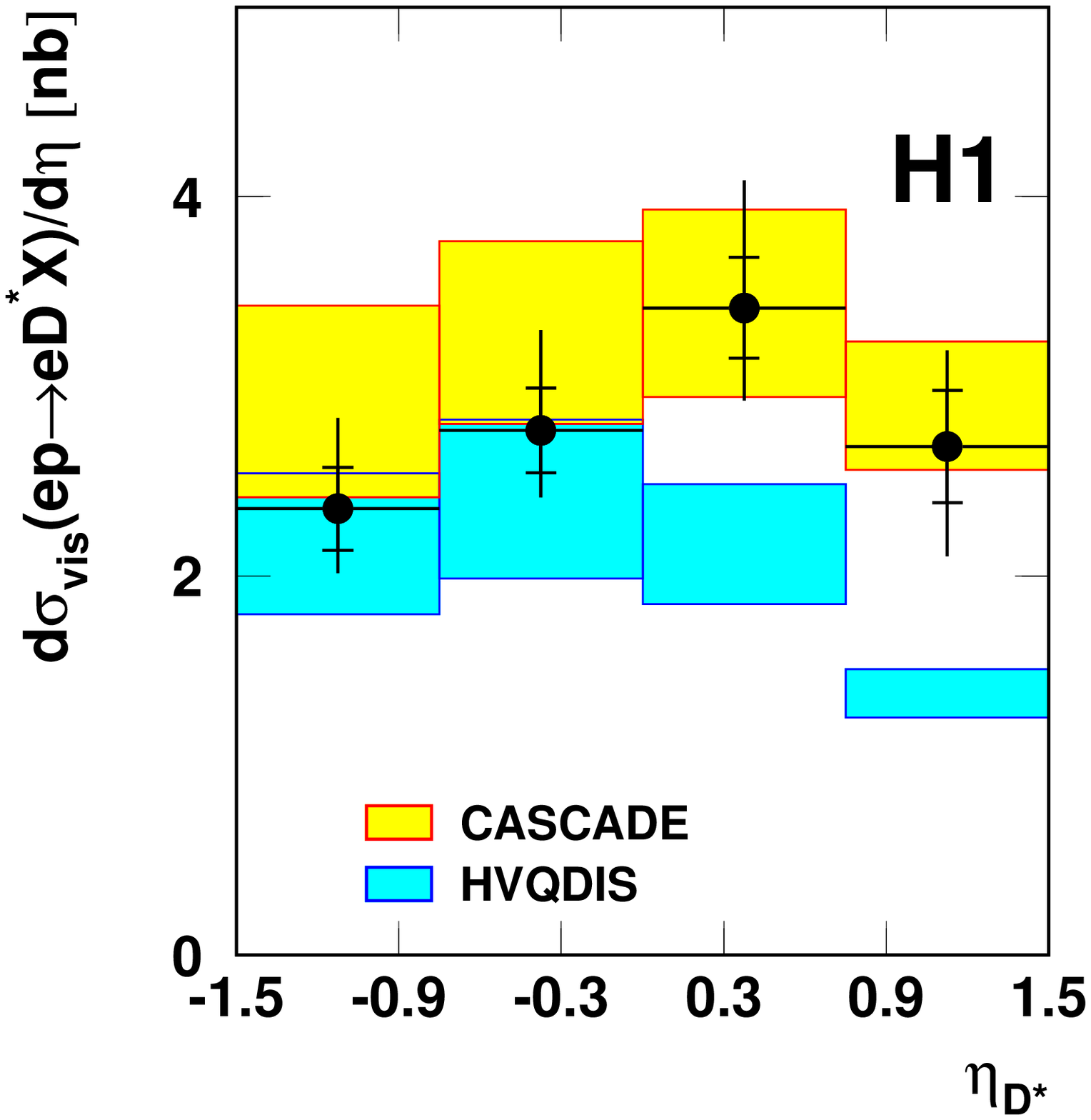,width=7.5cm}}
\put(1,-1){\epsfig{file=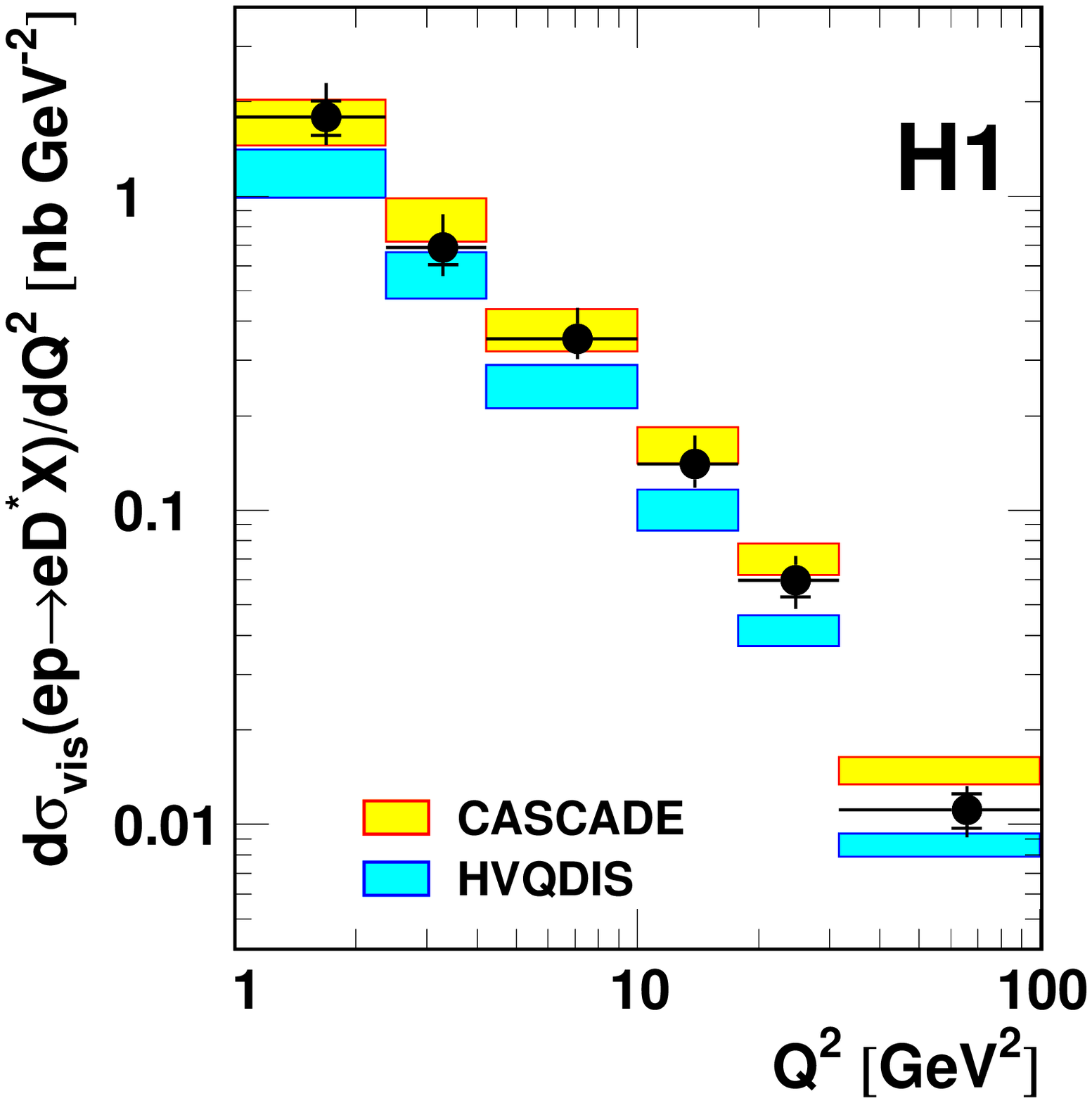,width=7.5cm}}
\put(8,-1){\epsfig{file=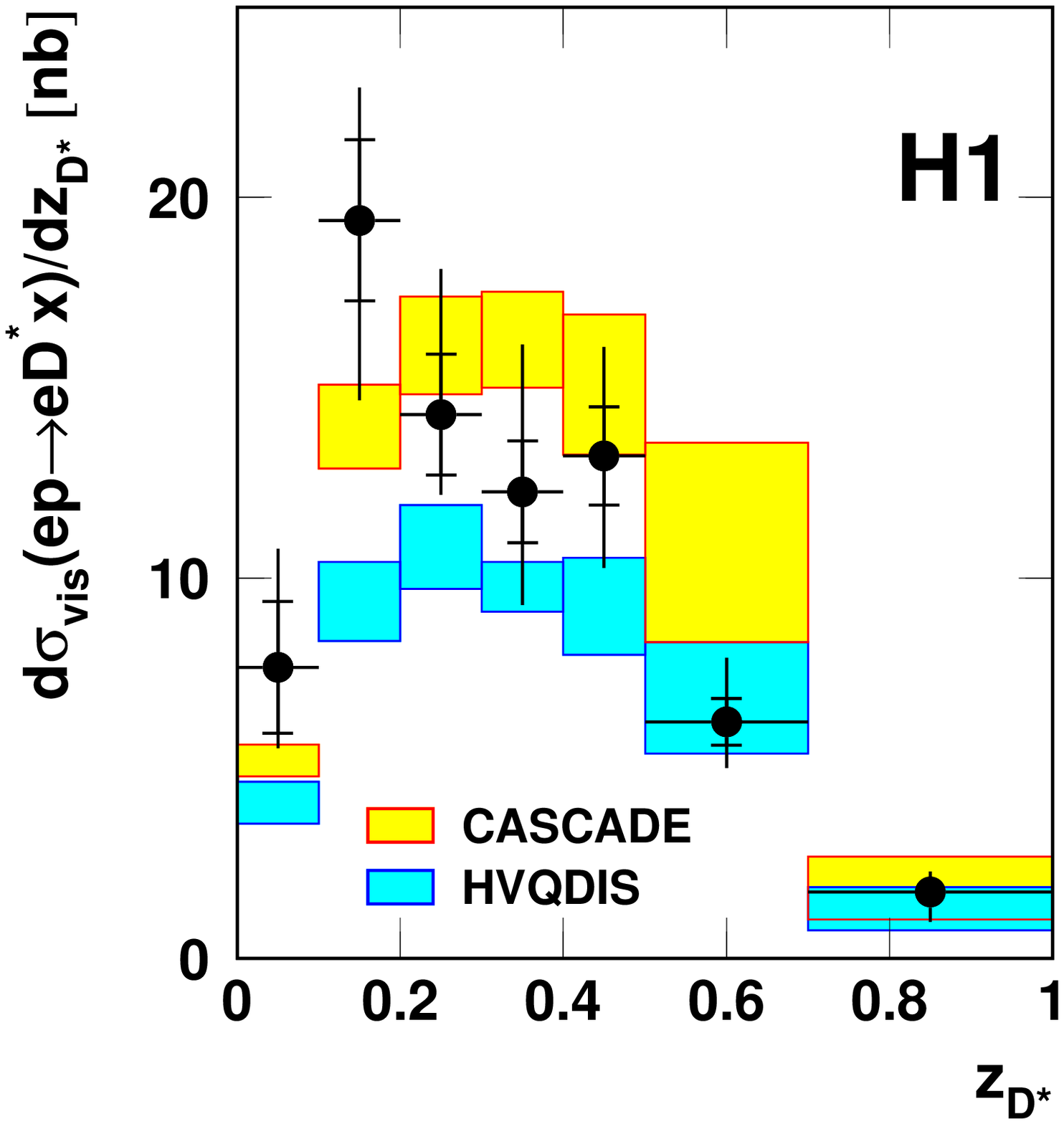,width=7.5cm}}
\end{picture}
\caption{\label{fig2}{Single
differential inclusive cross section $\sigma(ep \rightarrow eD^{*\pm} X)$
versus $W$, $x$, $Q^2$ and $p_{t\,D^*}$, $\eta_{D^*}$, $z_{D^*}$.
The inner and outer error bars correspond to the 
statistical and the total errors.  
The expectation of the NLO DGLAP calculation using HVQDIS with
GRV98-HO parton densities is indicated by the
lower shaded band. The upper shaded band is the expectation of the CCFM calculations based on the
CASCADE program with the initial gluon distribution fitted to the inclusive
$F_2$ data. 
The upper and lower bounds of both calculations correspond to ($m_c=1.3~\gev$,
$\epsilon_c=0.035$) and ($m_c=1.5~\gev$, $\epsilon_c=0.10$), respectively.
}}
\end{figure}

In Fig.~\ref{fig2} the inclusive single differential $D^{*\pm}$ cross 
sections in the visible region are shown as a function of the event 
variables $W$, $x$ and $Q^2$ and
as a function of the $D^{*\pm}$ observables $p_{t\,D^*}$, 
$\eta_{D^*}$ and the inelasticity 
$z_{D^*}={P\cdot p_{D^*}}/{P\cdot q}={(E-p_z)_{D^*}}/{2yE_e}$, where 
$P$, $q$ and $p_{D^*}$ denote the four-momenta of the incoming proton,
the exchanged photon and the observed $D^{*\pm}$ meson, respectively.
A bin by bin correction to account for QED radiation has been applied. 

Fig.~\ref{fig2} also includes the expectations from the HVQDIS program  
using the GRV98-HO parton density parameterization.
The renormalization scale 
and the factorization scale are set to $\mu=\sqrt{Q^2+4m_c^2}$.
The charm quark mass and the fragmentation parameter have been varied
from $m_c=1.3$~GeV and $\epsilon_c=0.035$ to 
$m_c=1.5$~GeV and $\epsilon_c=0.10$. The dark shaded band indicates the
uncertainties in the predictions due to these variations.
Although the predicted visible cross section is smaller
than experimentally observed, the agreement with the data in the shapes of the 
different single differential cross sections is reasonable. 
A significant difference is observed in the d$\sigma$/d$\eta$
cross section. For $\eta_{D^*}>0$ the measured $D^{*\pm}$ meson 
production cross section is  
larger than predicted by the 
calculation. Since in the boson gluon fusion process the forward region 
($\eta_{D^*}>0$) is correlated with small $z_{D^*}$ a similar discrepancy 
between data and theory is observed at small $z_{D^*}$.

A possible cause of this deviation could be the simplified grafting 
of fragmentation onto the HVQDIS program. This approach 
 does not account for the colour force between the charm quark and the proton 
remnant which is expected to result in a drag of the $D^{*\pm}$ meson from 
the original charm quark direction towards the proton direction.
To quantify this `beam drag effect' \cite{norrbin}, a mapping function from  
the $\eta_{c}$-$p_{t\,c}$ space to the $\eta_{D^*}$-$p_{t\,D^*}$ space 
has been constructed using the AROMA Monte Carlo program which includes 
such effects. This function
has then been used instead of  the Peterson fragmentation with transverse 
momentum smearing in the HVQDIS program. No significant change in the 
$\eta_{D^*}$ and $z_{D^*}$ distributions has been observed by this procedure 
compared to our original treatment of fragmentation. 
A better description of the  $\eta_{D^*}$ distribution is obtained, however,
when using the HERWIG program to extract the mapping function, at the expense 
of a 10-15\% reduction in the visible cross 
section prediction. It is therefore concluded that
the differences between the measurements and 
the predictions from the HVQDIS program can not be explained by the 
absence of colour drag effects in these calculations.  

Fig.~\ref{fig2} also presents 
the predictions of the CASCADE program with the same variations of 
the charm quark mass and the fragmentation parameter.
The expectations from the CASCADE program are found to agree better with the
data in general and especially in the positive $\eta$ region.

\begin{figure}[htb] 
\begin{center}
    \epsfig{file=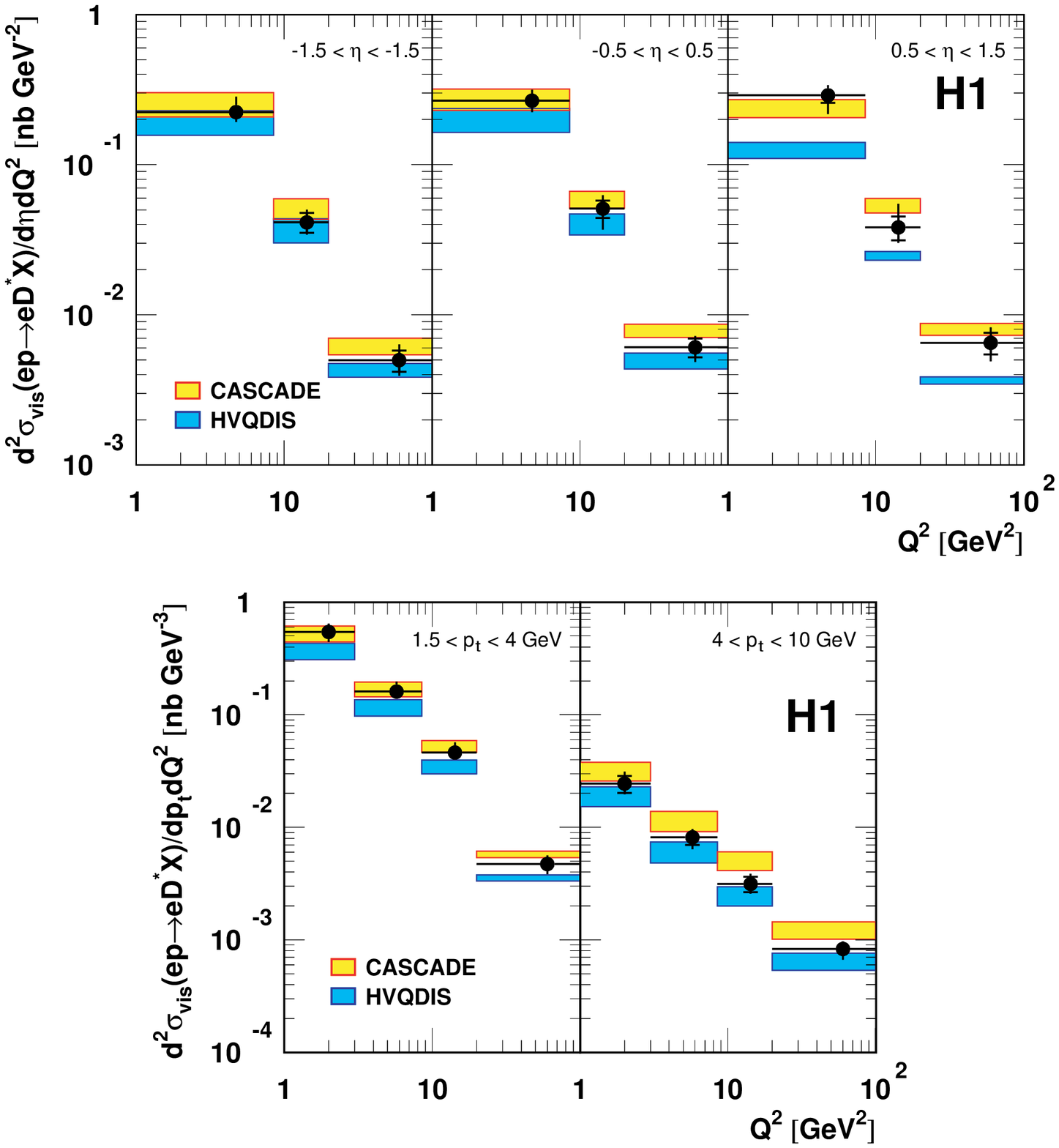,width=0.9\textwidth}
\caption{\label{fig4}{Double differential inclusive cross section 
${\rm d}^2\sigma/{\rm d}\eta{\rm d}Q^2$ and
${\rm d}^2\sigma/{\rm d}p_t{\rm d}Q^2$ in bins of $\eta_{D^*}$ and 
$p_{t\,D^*}$
(see figure \ref{fig2} for details).
}}
\end{center}  
\end{figure}

\begin{figure}[htb] 
\begin{center}
    \epsfig{file=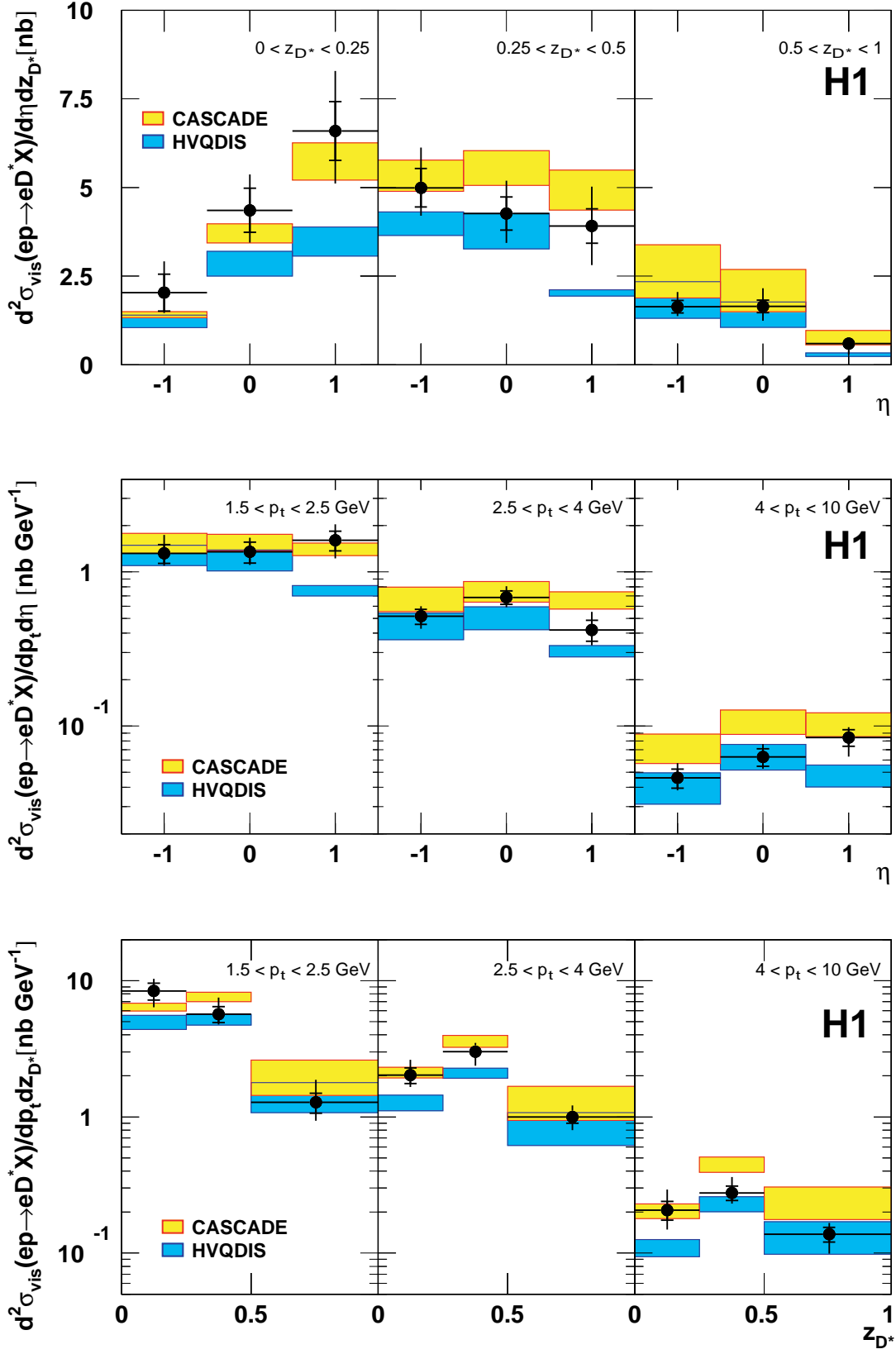,width=0.9\textwidth}
\caption{\label{fig5}{Double differential inclusive cross section 
${\rm d}^2\sigma/{\rm d}\eta{\rm d}z_{D^*}$ in bins of $z_{D^*}$ and
${\rm d}^2\sigma/{\rm d}p_t{\rm d}z_{D^*}$ and
${\rm d}^2\sigma/{\rm d}p_t{\rm d}\eta$ in bins of $p_{t\,D^*}$
(see figure \ref{fig2} for details).
}}
\end{center}  
\end{figure}

 In order to enable the study of correlations among the observables in 
$D^{*\pm}$ meson production, Figs.~\ref{fig4} and \ref{fig5} show the double 
differential inclusive $D^{*\pm}$ cross sections. 
It is evident that the excess observed in the data with respect to
the HVQDIS expectation at large pseudorapidities ($0.5<\eta_{D^*}<1.5$) 
is independent of $Q^2$ and is concentrated at small $p_{t\,D^*}$ 
and small $z_{D^*}$. 
It is especially in this phase space region where the CASCADE program
better represents the data. 

%
%
\section{Charm Contribution to the Proton Structure Function}
The charm contribution, $F^{c}_2(x,Q^2)$, to the proton structure function
is obtained by using the expression for the  
one photon exchange cross section for charm production
\begin{equation}
\displaystyle
\frac{d^2\sigma^{c}}{dxdQ^2}=\frac{2\pi\alpha_{em}^2}{Q^4x}
\left(1+\left(1-y\right)^2\right)\;F^{c}_2(x,Q^2)\;,
\end{equation}
where the contribution of the longitudinal structure function is neglected.
The visible inclusive $D^{*\pm}$ cross sections 
$\sigma_{\rm{vis}}^{\rm{exp}}(x,Q^2)$ in bins of $x$ and $Q^2$ are 
converted to a bin center corrected 
$F_2^{c~\rm{exp}}(\langle x\rangle,\langle Q^2\rangle)$ by the relation:
\begin{equation}
F_2^{c~\rm{exp}}(\langle x\rangle,\langle Q^2\rangle)=
\frac{\displaystyle \sigma_{\rm{vis}}^{\rm{exp}}(x,Q^2)}
{\displaystyle \sigma_{\rm{vis}}^{\rm{theo}}(x,Q^2)}\cdot
F_2^{c~\rm{theo}}(\langle x\rangle,\langle Q^2\rangle)~,
\label{f2cexp}
\end{equation}  
where $\sigma_{\rm{vis}}^{\rm{theo}}$ and $F_2^{c~\rm{theo}}$ are the 
theoretical  predictions from the model under consideration. The measured 
values of the visible cross sections, $\sigma_{\rm{vis}}^{\rm{exp}}(x,Q^2)$, 
are listed in Table \ref{sigvisxq2}. 
Following the same line as in previous publications 
\cite{h1f2c,zeusf2c,h1gluon} the HVQDIS program by Harris and Smith 
\cite{hvqdis}
and the program of Riemersma et al. \cite{riemersma} are used to calculate 
these quantities in the NLO DGLAP scheme. In the kinematic range of the 
current analysis
the beauty contribution to the proton structure, $F^b_2$, 
is expected to be of the order of 1 to 2\% of $F^c_2$ 
\cite{Daum:1996ec,Chuvakin:2000jm} and is therefore neglected~\footnote{If,
however, the beauty cross section turns out to be large
\cite{Sloan:2001hv}, its
contribution to $D^{*\pm}$ meson production may have to be
subtracted from the visible inclusive cross section prior to the determination
of $F_2^c$.}.
 
\begin{figure}[htb] 
\begin{picture}(17,18.5)
\put(1.8,9.5){\epsfig{file=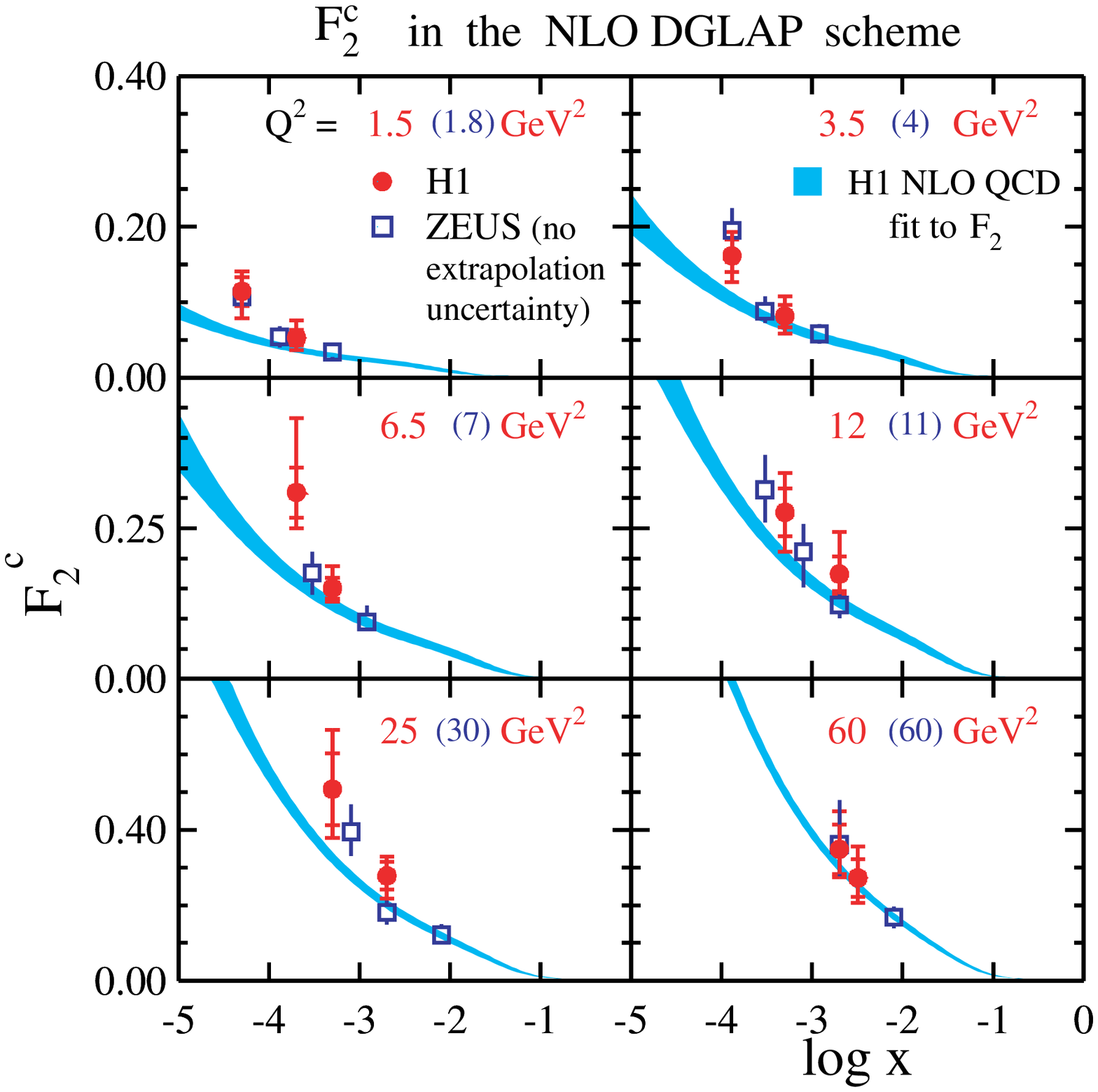,width=0.65\textwidth}}
\put(1.,8){{\Large (b)}\normalsize}
\put(1.,17.5){{\Large (a)}\normalsize}
\put(1.8,0.){\epsfig{file=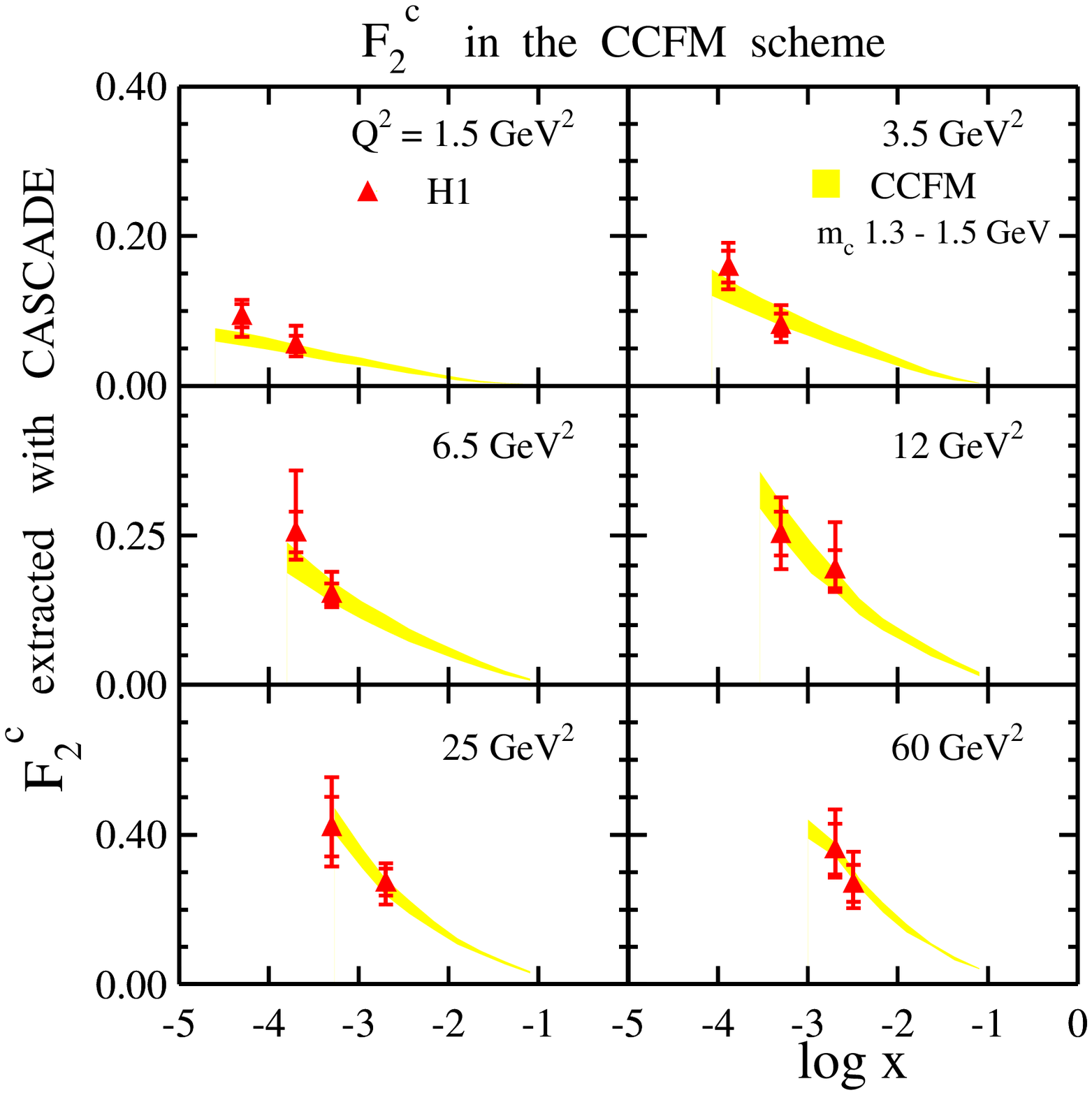,width=0.65\textwidth}}
\end{picture}
\caption{\label{fig6}{
$F_2^c$, as derived from the inclusive $D^{*\pm}$ meson analysis (a)
in the framework of NLO DGLAP and (b) in the framework of CCFM, both 
for $m_c=1.4~\gev$. The error bars on the H1 data points
refer to the statistical 
(inner) and the total (outer) error, respectively. In (a)
the shaded bands represent the predictions of $F_2^c$ 
from the H1 NLO DGLAP fit to the inclusive $F_2$ measurements
including all the uncertainties described in the text.
The dominant contribution arises from the uncertainty of $m_c$.
The ZEUS measurements~\cite{zeusf2c1} are shown for comparable
values of $Q^2$ indicated in parantheses
(see~\cite{zeusf2c1} for a discussion of the extrapolation uncertainties).
In (b) the bands represent the expectation of $F_2^c$
from the fit to the inclusive
$F_2$ in the CCFM scheme including only the variation of $m_c$ which 
in both
cases ranges from $1.3$ to $1.5~\gev$.}}
\end{figure}

In Fig.~\ref{fig6}a $F^{c}_2$ is shown as a function of $x$ for different 
values of $Q^2$ as extracted from the inclusive $D^{*\pm}$ 
cross sections using $m_c=1.4~\gev$. The systematic error on the data points
includes those described in Sec. \ref{difwq} as well as additional errors 
coming from the extrapolation in Equation~\ref{f2cexp}.
The bands show the predictions based on the gluon density
extracted by the H1 NLO DGLAP fit to the inclusive $F_2$ measurement 
\cite{h1nlo2000}. The width of each band reflects the total uncertainty of 
the prediction resulting from the uncertainties on this fit, 
thereby exploiting the full correlations arising from the constraints of
the inclusive $F_2$ measurement. The 
influence of all the individual sources of 
uncertainties considered in Ref. \cite{h1nlo2000}
for the determination of the gluon density and the 
strong coupling constant $\alpha_s$  has also been investigated here. 
The most relevant variations with respect to theoretical 
calculations are the variation of the strong coupling constant $\alpha_s$ 
in the range $0.113\le\alpha_s\le0.167$, of the factorization  
and renormalization scale $\mu$ in the range  
$0.5\cdot (Q^2+4m_c^2)\le\mu\le 2\cdot (Q^2+4m_c^2)$ and of the 
charm quark mass in the range $1.3\le m_c\le1.5\gev$. The
dominant sources of uncertainties are the experimental error on the $F_2$ 
measurement at very small $x$ and the insufficient knowledge of the charm 
quark mass in the range of the direct $F_2^c$ measurements.  
For the displayed bands the different contributions are added in quadrature. 
Fig.~\ref{fig6}a also includes the results
of \cite{zeusf2c1} for comparable
values of $Q^2$.
These measurements suggest a steeper rise at small $Q^2$ towards small $x$
than expected from the calculations based on the 
gluon density in the proton extracted from the inclusive $F_2$ measurement.
     
The extraction of $F^{c}_2$ according to Equation~\ref{f2cexp} is faced with an
intrinsic problem.  
The measurement covers about 30\% of the total phase space for 
charm production and the estimation of this acceptance fraction depends 
significantly on the underlying model. To be more explicit, two different 
calculations  may yield the same value for
$F_2^{c~\rm{theo}}(\langle x\rangle,\langle Q^2\rangle)$ but may have 
different acceptances. They may then predict for a bin in $x$ and $Q^2$
different cross sections $\sigma_{\rm{vis}}^{\rm{theo}}(x,Q^2)$ with 
consequent different values for $F_2^{c~\rm{exp}}$. 
To investigate the model dependence $F_2^{c~\rm{exp}}$ is also determined
using the CASCADE program with $m_c=1.4~\gev$ and the results are shown in
Fig.~\ref{fig6}b.
The figure includes also the prediction according to the CCFM evolution.
Here the bands indicate the uncertainty on this prediction due to the 
variation of the charm quark mass.
The comparison of Figs.~\ref{fig6}a and \ref{fig6}b reveals a steeper 
rise in
the predicted charm contribution to the proton structure function at small $x$
in the CCFM evolution than obtained by the NLO DGLAP 
evolution. Using the acceptances and efficiencies calculated from the 
CASCADE program 
the measured values of $F^{c}_2$ are found to be systematically smaller than 
those determined with the HVQDIS program. 
The largest differences (up to $\approx 20$\%) are observed at small $x$ 
values.

\begin{figure}[htb]
    \begin{center}
\epsfig{file=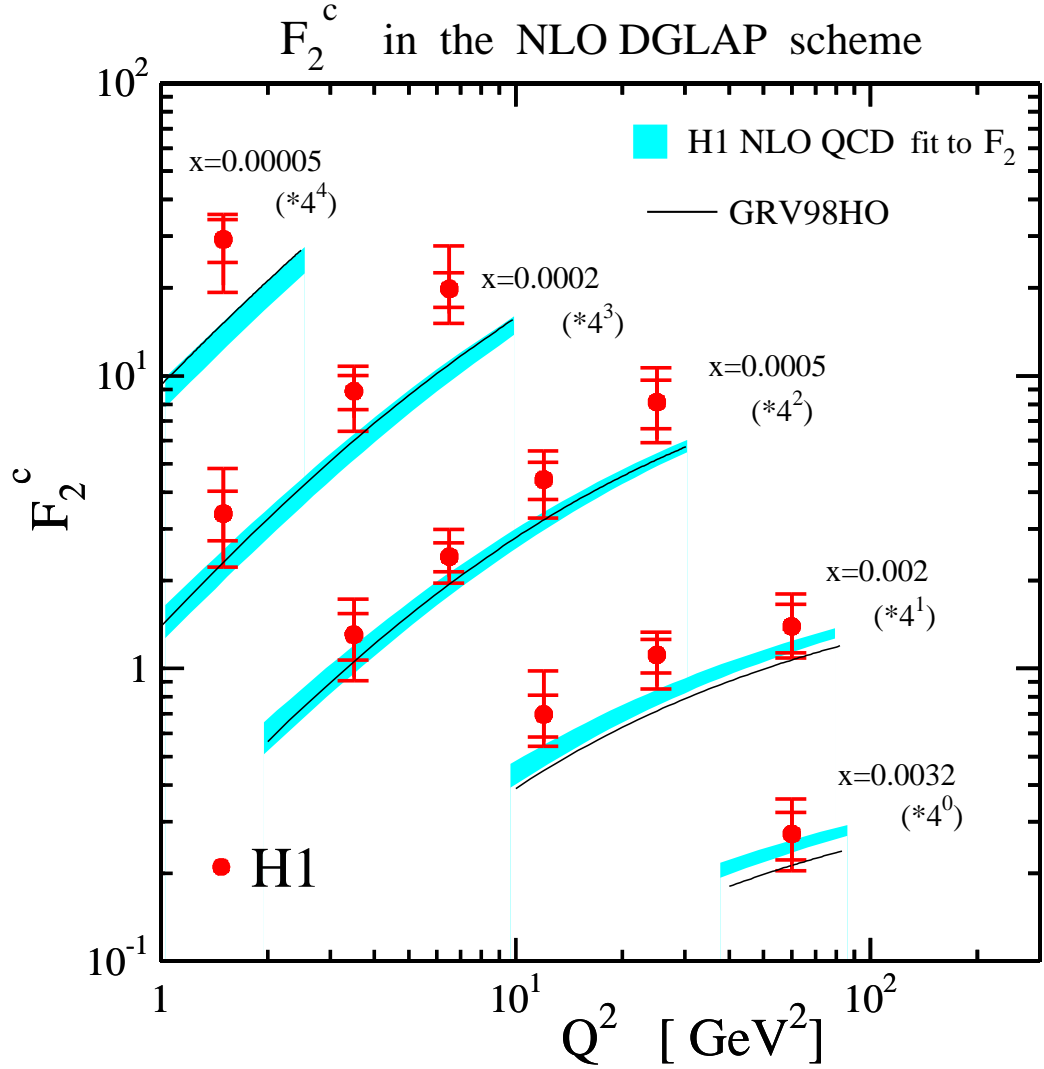,width=0.95\textwidth}
\caption{\label{fig8} $F_2^c$ as derived from the inclusive $D^{*\pm}$ meson 
production cross
section as a function of $Q^2$ for different values of $x$.
The error bars refer to the statistical 
(inner) and the total (outer) error, respectively.
The shaded bands represent the predictions of the NLO DGLAP evolution 
based on the parton densities in the proton obtained by the fit to the 
inclusive $F_2$ for $m_c=1.4~\gev$ including all uncertainties described 
in the text.
The black lines show the predictions in the NLO DGLAP scheme  for 
$m_c=1.4~\gev$ using the gluon density from GRV98-HO. }
    \end{center}
\end{figure}

In Fig.~\ref{fig8} 
$F^{c}_2$ is shown as a function of $Q^2$ for different values of $x$ using 
the acceptances as calculated with HVQDIS. As in Fig.~\ref{fig6} the bands 
indicate the full uncertainty in the DGLAP NLO predictions for  a central 
value of the charm quark mass of $1.4\gev$
using the gluon density extracted from the fit to the inclusive $F_2$ 
measurement. The full line shows the DGLAP NLO prediction using 
the gluon density from GRV98-HO. Taking into account the different data sets 
used for the determinations of the gluon densities, the agreement of the 
different calculations is reasonable. 
The data show a steep rise of $F_2^c$ with $Q^2$. The slope, 
$\partial F_2^c/\partial \ln Q^2$,
contributes roughly half of the slope of the inclusive structure
function, $\partial F_2/\partial \ln Q^2$, measured at the central $Q^2$ 
of each $x$ bin
in \cite{h1nlo2000} in the range $0.0002<x<0.002$. This steep rise is
reasonably well reproduced by the NLO DGLAP calculations.

\begin{figure}[htb] 
    \begin{center}
    \epsfig{file=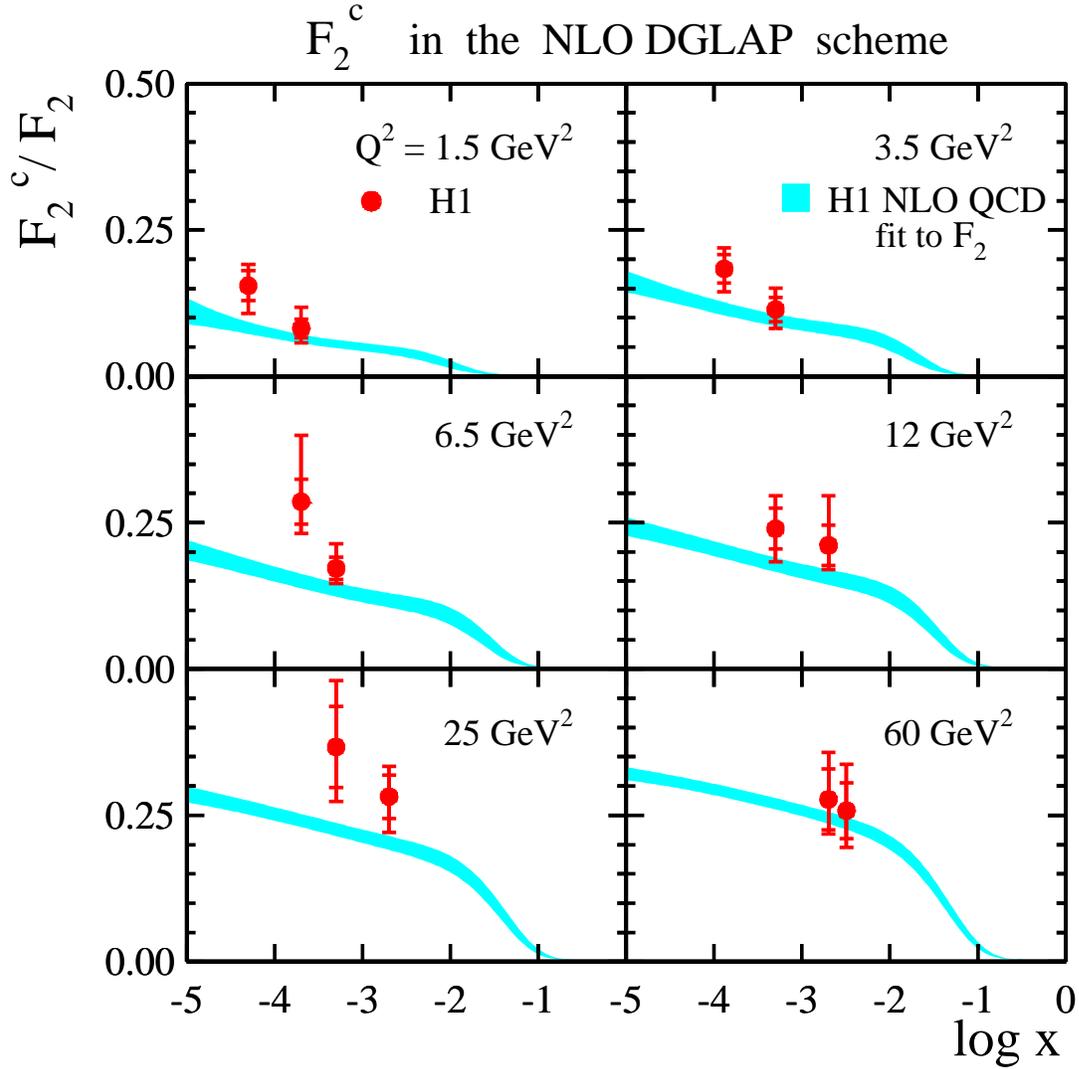,width=0.95\textwidth}
\caption{\label{fig9}{The ratio of 
$F_2^c$ over $F_2$ as derived from the inclusive $D^{*\pm}$ meson analysis
as a function of $x$ for different values of $Q^2$.
 The error bars refer to the statistical 
(inner) and the total (outer) error, respectively.
The shaded bands represent the predictions of the NLO DGLAP evolution 
based on the parton densities in the proton obtained by the fit to the 
inclusive $F_2$ for a central charm quark mass of  
$1.4~\gev$ including all uncertainties.}}
    \end{center}
\end{figure}

In Fig.~\ref{fig9} the ratio of $F_2^c$ to the inclusive $F_2$
\cite{h1nlo2000} is shown as a function of $x$ for different values of $Q^2$. 
The contribution of charm production to the total $F_2$ rises from about 10 \%
at $Q^2$~=~1.5~GeV$^2$ and $x\approx 10^{-4}$ to more than 25 \% at 
$Q^2~\ge~25\gev^2$ and $x\ge 5\cdot10^{-4}$. Although this behaviour agrees 
with expectation, at small $x$ the measured ratio of $F_2^c/F_2$ is larger 
than predicted in the NLO DGLAP scheme.

\section{Conclusions}

New measurements of differential cross sections for inclusive
$D^{*\pm}$ production in deep-inelastic $ep$ scattering are presented. 
These are compared with predictions based on both NLO DGLAP and CCFM
formalisms, the former made using the HVQDIS program~\cite{hvqdis} and
the latter using the CASCADE model~\cite{cascade}.  The predictions
made using DGLAP formalism tend to undershoot the data, particularly
for small $D^*$ transverse momenta, $p_{t\,D^*}$, and positive $D^*$
pseudorapidities, $\eta_{D^*}$.  The expectations of the CCFM based
model are in better agreement with the data.

Extrapolation of the visible $D^*$ cross section to the full
$p_{tD^*}$ and $\eta_{D^*}$ phase space allows extraction of $F_2^c$,
the contribution of charm to the proton structure function $F_2$. 
These extrapolations are seen to depend on the formalism used: that
based on the NLO DGLAP formalism typically produces a larger result
for $F_2^c$ than that made using the CCFM approach.  Both results are
presented to allow consistent comparisons using either formalism. 
The kinematic range presented has been extended to lower $Q^2$ than
shown in the previous H1 study, namely $Q^2 = 1\,$GeV$^2$, and thereby
to lower $x$.  The $F_2^c$ measurements show large scaling violations
and a steep rise of $F_2^c$ with decreasing $x$.  This rise tends to
be steeper than expected from the NLO DGLAP calculations, but agrees
well with the CCFM based expectations.  Both approaches for the
extraction of $F_2^c$ show that the contribution of charm production
to $F_2$ exceeds $25\%$ for $Q^2 > 25\,$GeV$^2$.

\section*{Acknowledgments}

We are grateful to the HERA machine group whose outstanding
efforts have made and continue to make this experiment possible. 
We thank
the engineers and technicians for their work in constructing and now
maintaining the H1 detector, our funding agencies for 
financial support, the
DESY technical staff for continual assistance 
and the DESY directorate for the
hospitality which they extend to the non DESY 
members of the collaboration.

\begin{table}[hb]
\begin{center}
\begin{tabular}{lrr}\hline\noalign{\smallskip}
Experimental Systematic Uncertainties\cr
\noalign{\smallskip}\hline\noalign{\smallskip}
Trigger efficiency&&$\pm$~0.02\cr\noalign{\smallskip}
Track detector efficiency&$+$0.075&$-$0.034
\cr\noalign{\smallskip}
d$E$/d$x$ measurement&&$\pm$~0.03\cr\noalign{\smallskip}
$p_t(\pi_{\rm{slow}})$-cut&&$+$0.025\cr\noalign{\smallskip}
Background shape&&$\pm$~0.05\cr\noalign{\smallskip}
$D^0$ Mass resolution&$+$0.04&$-$0.025\cr\noalign{\smallskip}
Reflections&&$\pm$~0.015\cr\noalign{\smallskip}
Event kinematics&&$\pm$0.04\cr\noalign{\smallskip}
Luminosity measurement&&$\pm$~0.015\cr
$\gamma p$ contribution&&$<$0.004\cr\noalign{\smallskip}
Branching ratio&&$\pm$~0.025\cr
\noalign{\smallskip}\hline 
\noalign{\smallskip}
&$+$0.12&$-$0.09\cr\noalign{\smallskip}
\hline\noalign{\smallskip}
\end{tabular}
\caption{Summary of the fractional experimental systematic uncertainties of the
inclusive $D^{*\pm}$ meson cross section.}
\label{expsys}
\end{center}
\end{table}

\begin{table}
\begin{center}
\begin{tabular}{lrr}\hline\noalign{\smallskip}
Model Uncertainties&&\cr
\noalign{\smallskip}\hline\noalign{\smallskip}
Fragmentation model
&$+$0.035&$-$0.07\cr\noalign{\smallskip}
Charm quark mass&$+$0.07&$-$0.02\cr\noalign{\smallskip}
Scale $\mu=\sqrt{Q^2+4m_c^2}$&&$-$0.015\cr\noalign{\smallskip}
QED radiation&&$-$0.025\cr\noalign{\smallskip}
\noalign{\smallskip}
\hline 
\noalign{\smallskip}
&&$\pm$0.08
\cr\noalign{\smallskip}\hline
\noalign{\smallskip}
\end{tabular}
\caption{Summary of the fractional model dependent uncertainties of the
 inclusive $D^{*\pm}$ meson cross sections.}
\label{modelsys}
\end{center}
\end{table}

\begin{table}
\begin{center}
{\footnotesize
\begin{tabular}{lrcrccc}\hline\noalign{\smallskip}
$\log\frac{\,Q^2}{\gev^2}$&$\langle Q^2\rangle$
&$\log{(x)}$&$\langle x\rangle$&$\sigma_{vis}$&$F_2^c$&$F_2^c$ \cr
&$[\gev^2]$&&$[10^{-3}]$&[nb]&(DGLAP)&(CCFM)\cr
\noalign{\smallskip}
\hline\noalign{\smallskip}
$0$&$1.5$&-$4.8$~to~-$4.2$&$0.05$&$1.11\pm0.19~{+0.15\atop-0.28}$&$0.114\pm0.019~{+0.019\atop-0.030}$&$0.094\pm0.016~{+0.014\atop-0.024}$\cr
\noalign{\smallskip}
\noalign{\smallskip}
to~$0.375$&&-$4.2$~to~-$3.0$&$0.20$&$1.44\pm0.28~{+0.55\atop-0.34}$&$0.053\pm0.010~{+0.021\atop-0.013}$&$0.056\pm0.011~{+0.022\atop-0.013}$\cr
\noalign{\smallskip}
\hline\noalign{\smallskip}
$0.375$&$3.5$&-$4.6$~to~-$3.6$&$0.13$&$0.95\pm0.13~{+0.14\atop-0.13}$&$0.161\pm0.022~{+0.023\atop-0.027}$&$0.159\pm0.021~{+0.024\atop-0.022}$\cr
\noalign{\smallskip}
\noalign{\smallskip}
to~$0.625$&&-$3.6$~to~-$3.0$&$0.50$&$0.41\pm0.07~{+0.11\atop-0.09}$&$0.082\pm0.015~{+0.022\atop-0.018}$&$0.082\pm0.015~{+0.022\atop-0.018}$\cr
\noalign{\smallskip}
\hline\noalign{\smallskip}
$0.625$&$6.5$&-$4.0$~to~-$3.6$&$0.20$&$0.72\pm0.10~{+0.27\atop-0.08}$&$0.309\pm0.041~{+0.116\atop-0.042}$&$0.256\pm0.034~{+0.097\atop-0.031}$\cr
\noalign{\smallskip}
\noalign{\smallskip}
to~$1$&&-$3.6$~to~-$3.0$&$0.50$&$1.02\pm0.11~{+0.21\atop-0.10}$&$0.151\pm0.017~{+0.032\atop-0.015}$&$0.153\pm0.017~{+0.032\atop-0.015}$\cr
\noalign{\smallskip}
\hline\noalign{\smallskip}
$1$&$12$&-$3.6$~to~-$3.0$&$0.50$&$0.59\pm0.09~{+0.11\atop-0.11}$&$0.276\pm0.040~{+0.052\atop-0.052}$&$0.253\pm0.037~{+0.048\atop-0.047}$\cr
\noalign{\smallskip}
\noalign{\smallskip}
to~$1.25$&&-$3.0$~to~-$2.0$&$2.00$&$0.43\pm0.07~{+0.16\atop-0.04}$&$0.174\pm0.029~{+0.064\atop-0.018}$&$0.194\pm0.032~{+0.071\atop-0.021}$\cr
\noalign{\smallskip}
\hline\noalign{\smallskip}
$1.25$&$25$&-$3.6$~to~-$3.0$&$0.50$&$0.33\pm0.06~{+0.08\atop-0.06}$&$0.508\pm0.096~{+0.125\atop-0.087}$&$0.422\pm0.079~{+0.105\atop-0.071}$\cr
\noalign{\smallskip}
\noalign{\smallskip}
to~$1.5$&&-$3.0$~to~-$2.0$&$2.00$&$0.54\pm0.07~{+0.07\atop-0.09}$&$0.278\pm0.037~{+0.035\atop-0.048}$&$0.273\pm0.036~{+0.036\atop-0.047}$\cr
\noalign{\smallskip}
\hline\noalign{\smallskip}
$1.5$&$60$&-$3.0$~to~-$2.6$&$2.00$&$0.31\pm0.06~{+0.07\atop-0.03}$&$0.348\pm0.066~{+0.077\atop-0.034}$&$0.362\pm0.068~{+0.081\atop-0.035}$\cr
\noalign{\smallskip}
\noalign{\smallskip}
to~$2$&&-$2.6$~to~-$2.0$&$3.16$&$0.35\pm0.06~{+0.09\atop-0.06}$&$0.272\pm0.050~{+0.067\atop-0.044}$&$0.270\pm0.050~{+0.069\atop-0.043}$\cr
\noalign{\smallskip}
\hline
\end{tabular}
}
\caption{Inclusive $D^{*\pm}$ cross section $\sigma_{vis}$
 in bins in $x$ and $Q^2$
for the visible range $|\eta_{D^*}|<1.5$ and $p_{t\,D^*}>1.5$~GeV as 
extracted in the DGLAP scheme. 
 The values given in the columns denoted by 
$\langle Q^2\rangle$ and $\langle x\rangle$ are the bin centres at which the
values of $F_2^c$ are given. The values of $F_2^c$ extracted in both
the DGLAP and CCFM schemes are also presented.
} 
\label{sigvisxq2}
\end{center}
\end{table}

\unitlength1cm

\end{document}